\documentstyle{mn}
\input{psfig}

\title[The Evolution of Rapid~Burster Outbursts]{The Evolution of Rapid~Burster Outbursts}

\author[R. Guerriero et al.]{R.~Guerriero,$^1$ D. W.~Fox,$^1$
 J.~Kommers,$^1$ W. H. G.~Lewin,$^1$ R.~Rutledge,$^2$ C. B.~Moore,$^3$
\newauthor  E.~Morgan,$^1$ J.~Van Paradijs,$^{4,5}$ M.~Van der Klis,$^{5,6}$
 L.~Bildsten,$^6$ T.~Dotani$^7$\\ 
$^1$MIT Center for Space Research, 77 Massachusetts Ave \#37-627,
 Cambridge, MA 02139-4307, USA\\
$^2$Dept.\ of Astronomy, Caltech MC 220-47, Pasadena, CA 91125, USA\\
$^3$Kapteyn Astronomical Institute, Postbus 800, 9700 AV Groningen,
 The Netherlands\\
$^4$Univ.\ of Alabama in Huntsville, Huntsville, AL 35812, USA\\
$^5$Astronomical Institute `Anton Pannekoek' and Center for
 High-Energy Physics, Kruislaan 403, 1098 SJ Amsterdam, The
 Netherlands\\
$^6$Depts.\ of Physics and Astronomy, University of California,
 Berkeley, Berkeley, CA 94720, USA\\ 
$^7$Institute of Space and Astronautical Science, Sagamihara, Japan }

\date{Received date.  Accepted date.}

\begin{document}

\newcommand{\qq}{\mbox{``}}
\newcommand{\chinu}{\mbox{$\chi_\nu^2$}}
\newcommand{\bs}{\bigskip}
\newcommand{\ms}{\medskip}
\newcommand{\rxte}{\textit{RXTE}}
\newcommand{\rxtez}{\textit{RXTE\/}}
\newcommand{\exo}{\textit{EXOSAT\/}}
\newcommand{\hak}{\textit{Hakucho\/}}
\newcommand{\tenma}{\textit{Tenma\/}}
\newcommand{\ging}{\textit{Ginga\/}}
\newcommand{\sas}{\textit{SAS-3\/}}
\newcommand{\ariel}{\textit{Ariel V\/}}
\newcommand{\heao}{\textit{HEAO I\/}}
\newcommand{\ein}{\textit{Einstein\/}}
\newcommand{\uhur}{\textit{Uhuru\/}}
\newcommand{\cop}{\textit{Copernicus\/}}
\newcommand{\groj}{GRO~J1744$-$28}
\newcommand{\mxbrb}{MXB~1730$-$335}
\newcommand{\slowb}{4U~1728$-$34}
\newcommand{\ergsec}{\mbox{erg s$^{-1}$}}

\maketitle

\begin{abstract}
We describe the evolutionary progression of an outburst of the Rapid
Burster.  Four outbursts have been observed with the \textit{Rossi
X-Ray Timing Explorer\/} between February 1996 and May 1998, and our
observations are consistent with a standard evolution over the course
of each.  An outburst can be divided into two distinct phases: Phase I
is dominated by type~I bursts, with a strong persistent emission
component; it lasts for 15--20 days.  Phase II is characterized by
type~II bursts, which occur in a variety of patterns.  The light
curves of time-averaged luminosity for the outbursts show some
evidence for reflares, similar to those seen in soft X-ray transients.
The average recurrence time for Rapid Burster outbursts during this
period has been 218 days, in contrast with an average $\sim$180 day
recurrence period observed during 1976--1983.

\end{abstract} 


\section{Introduction}
The Rapid Burster (\mxbrb, or RB hereafter; Lewin et al.\ 1976),
discovered in 1976, is a unique recurrent transient low-mass X-ray
binary (LMXB). It is located at a distance of approximately 8~kpc
(Ortolani, Bica, \& Barbuy 1996) in the highly reddened globular
cluster Liller~1 (Liller 1977).  The RB is the only known LMXB to
produce both type~I and type~II X-ray bursts (Hoffman, Marshall, \&
Lewin 1978a).

Although the RB has been studied for over twenty years, it is still
not clear why the RB, and only the RB, emits both type~I and type~II
X-ray bursts. Type~I bursts are due to a thermonuclear flash of
accreted material on the surface of a neutron star, and are
characterized by a distinct spectral softening during burst decay.  Of
the $\sim$125 LMXBs known, at least 43 are type~I burst sources (Van
Paradijs 1995).  Type~II bursts are due to spasmodic accretion -- the
release of gravitational potential energy -- presumably resulting from
an accretion disk instability; the spectrum of these bursts shows
little evolution during the burst.  The duration of type~II bursts can
range from 680 seconds (the longest type~II burst observed to date)
down to 4 seconds.  The behavior of type~II bursts is like that of a
relaxation oscillator: the type~II burst fluence $E$ is roughly
proportional to the time interval, $\Delta t$, to the following burst
(the `$E$-$\Delta t$' relation:~Lewin et al.\ 1976).  The type~II
burst luminosities at burst maximum range from
$\sim$4$\times$10$^{37}$ to $\sim$3$\times$10$^{38}$ \ergsec\ (Lewin,
Van Paradijs \& Taam, 1993; henceforth LVT93).  \groj\ is the only
other LMXB known to emit repetitive type~II bursts (Kouveliotou et
al. 1996; Lewin et al. 1996a; Kommers et al. 1997).  LVT93 provide a
comprehensive review of type~I and type~II X-ray bursts and the Rapid
Burster.

The pattern of type~I and type~II bursts, and the shape of the type~II
bursts themselves, have been observed to vary widely during a single
outburst.  At times, the RB emits only type~I bursts with strong
persistent emission (PE), behaving like a ``normal'' LMXB.  At other
times, type~II bursts, occurring in a wide variety of forms and
patterns, with or without substantial PE, are observed.  When the RB
is in a rapid bursting mode, it can emit thousands of rapid type~II
bursts per day, with little or no PE present.  Short type~II bursts
typically exhibit a timescale-invariant profile, with multiple peaks
(`ringing') during burst decay (Tawara et al.\ 1985).  Bursts longer
than about 35 seconds, on the other hand, are `flat-topped' in shape
(Lewin et al.\ 1976; Kunieda et al.\ 1984a; Stella et al. 1988a; Tan
et al.\ 1991).  An evolution from type~I to type~II bursting behavior
was observed previously during the August 1983 outburst (Kunieda et
al.\ 1984a; Barr et al.\ 1987)

Quasi-periodic oscillations (QPO) in the 2--8~Hz frequency range are
regularly seen in type~II bursts from the RB, and occasionally in the
PE (Tawara et al. 1982; Stella et al. 1988a,b; Dotani et al. 1990;
Lubin et al. 1991; Rutledge et al. 1995).  QPO are also present in
many RB type~II bursts in the form of the ringing observed during
burst decay (LVT93).  QPO have not been observed in any type~I bursts
from the RB.  There are strong 0.04~Hz QPO present in the PE after
some long type~II bursts (Lubin et al. 1992b).

\begin{figure}
\begin{center}
~\psfig{figure=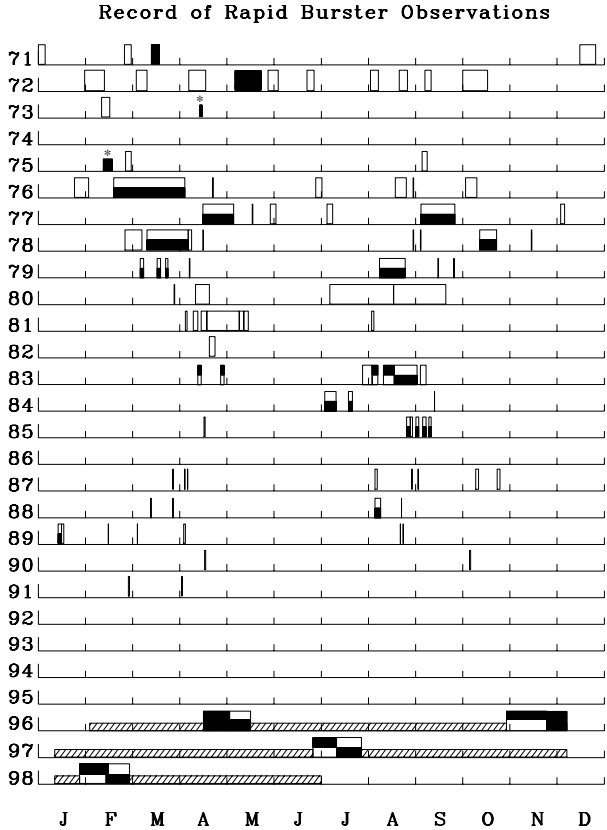,bbllx=43pt,bblly=35pt,bburx=565pt,bbury=760pt,width=3.2in}~
\caption{Record of observations of the Rapid Burster (1971-1998).
Filled boxes indicate observations that occurred when the RB was
active (a filled top half corresponds to Phase I, a filled bottom half
to Phase II, and a completely filled box indicates that the phase
could not be determined; see text).  Open boxes indicate observations
in which the RB was inactive.  Single lines indicate observations of
less than one day in which the RB was inactive.  The observations
marked with ``*'' indicate periods in which the source may have been
active, but this is uncertain (White et al., 1978).  Since February,
1996, the RB has been monitored for 11 months of the year with the
\rxtez\ ASM (slashed boxes).  This figure is an extension of similar
figures from Lewin and Joss (1983), and LVT93.}
\label{fig:cov}
\end{center} 
\end{figure}

The outbursts of the Rapid Burster have long been known to recur every
6--8 months, based on observed outbursts (LVT93).  Years at a time
have passed without any positive detections; however, monitoring of
the source has historically been sporadic (Figure~\ref{fig:cov}).
This changed in February 1996, when daily coverage of the RB (for 11
months of the year) began with the \textit{Rossi X-ray Timing
Explorer\/} (\rxtez) All-Sky Monitor (ASM) (Levine et al. 1996).

% This is now Figure 1b

Since then, we have observed four complete outbursts with {\it RXTE}.
These outbursts have recurred at intervals of 200, 217, and 238 days.
These outbursts show a nearly identical global evolutionary pattern:
all four outbursts evolved, on the same timescale, from an initial
phase dominated by type~I bursts to a type~II burst-dominated phase.
This evolutionary progression may provide some insight into the unique
behavior of the RB.

Rutledge et al. (1998) have observed a radio source at 4.5/8.4 GHz
whose strength is correlated with the X-ray emission from the RB as
measured by the \rxtez\ ASM.  They have proposed that this radio
emission comes from the RB, even though the $\sim$1$^{\prime\prime}$
position of the radio source lies well outside the 2$\sigma$ error
circle for the RB obtained with {\it Einstein\/} in 1984 (Grindlay et
al. 1984; Moore et al.\ 1999).

We report here on our analysis of data taken with \rxte\ during the
May 1996, November 1996, June 1997, and January 1998 outbursts of the
Rapid Burster (Lewin et al. 1996b,c; Guerriero et al. 1997, 1998).  In
Section~\ref{sec:obs} we summarize our observations; in
Section~\ref{sec:evolution} we present the general evolution of a RB
outburst; in Section~\ref{sec:historical} we compare our results to
past observations of the RB with other satellites, and in
Section~\ref{sec:discuss} we give some possible interpretations of our
findings.

\section{Observations and Analysis}
\label{sec:obs}
The RB was observed with \rxtez\ on 31 separate occasions from May
3--13, 1996, November 6--17, 1996, June 26--July 30, 1997, and January
30--February 19, 1998 (Table~\ref{tbl:obs}).  The total observing time
during these periods was 7.5~ksec, 12.6~ksec, 35.0~ksec, and
44.8~ksec, respectively.
\begin{table*}
\begin{minipage}{5.75in}
  \centering
\caption{\small \rxtez\ Observations of the Rapid Burster from
  1996--1998.  Day of outburst is calculated by considering the first
  $>$3$\sigma$ detection in the ASM to begin ``day 1''.  Counts refer
  to the \rxtez\ PCA (1 count s$^{-1}$ $\simeq$ 3$\times$10$^{-12}$
  erg cm$^{-2}$ s$^{-1}$; 2--20 keV).  Type~I burst durations were
  calculated by considering the end of the burst to be the point at
  which the excess burst flux returned to 10 per cent of its peak
  value.  The count rates for all bursts and persistent emission
  during offset pointings have been corrected for aspect (factor of
  2.5 for the RB).}
\label{tbl:obs}
  \begin{tabular}{rccc@{\hspace{.1in}}r@{\hspace{.1in}}c@{\hspace{.1in}}l@{\hspace{.1in}}cc}
\multicolumn{3}{c}{} & 
\multicolumn{1}{c}{Obs} &
\multicolumn{4}{c}{Bursts} & 
\multicolumn{1}{c}{Average RB}\\ \cline{5-8}
\multicolumn{2}{c}{Obs Start} & Day of & Duration & 
\multicolumn{3}{c}{Number} & Duration & PE Level\\
\multicolumn{2}{c}{(UT)} & Outburst & (ksec) & 
\multicolumn{1}{r}{Type I} & 
\multicolumn{2}{l}{Type II} & (s) & (cts s$^{-1}$) \\ \hline
1996 May 03 & 15:25 & 21 & 1.5 & 0~\, & + & 15 & 9--16 & ~400 \\
         07 & 15:37 & 25 & 3.0 & 0~\, & + & 8  & 8--17 & ~250 \\
         13 & 12:19 & 31 & 3.0 & 0~\, & + & 0 & n/a & ~180 \\
\\
1996 Nov 06 & 19:39 & ~8 & 2.6 & 1~\, & + & 0 & 250 & 2270 \\ 
     ~~~~09 & 00:31 & 11 & 2.7 & 1~\, & + & 0 & 200 & 1840 \\
     ~~~~10 & 16:34 & 12 & 3.0 & 0~\, & + & 0 & n/a & 1370\\
     ~~~~11 & 21:23 & 13 & 1.8 & 1~\, & + & 0 & 30 & 1590 \\
     ~~~~17 & 00:41 & 19 & 2.5 & 0~\, & + & 0 & n/a & 1090 \\
\\
1997 Jun 26 & 04:36 & ~2 & 1.6 & 3~\, & + & 0 & 50--60 & 4630 \\
     ~~~~26 & 08:01 & ~2 & 2.1 & 4~\, & + & 0 & 100--200 & 4630 \\
     ~~~~27 & 17:45 & ~3 & 3.0 & 2* & + & 0 & 150 & 4750 \\
     ~~~~29 & 06:26 & ~5 & 2.0 & 1~\, & + & 0 & 250 & 3320 \\
     ~~~~29 & 17:53 & ~5 & 2.6 & 2~\, & + & 0 & 150--180 & 3320 \\
     Jul~02 & 02:59 & ~8 & 1.4 & 0~\, & + & 0 & n/a & ~~2400$^{\dag}$ \\
     ~~~~03 & 02:59 & ~9 & 1.3 & 0~\, & + & 0 & n/a & ~~2400$^{\dag}$ \\
     ~~~~07 & 13:00 & 13 & 2.5 & 1~\, & + & 0 & 130--150 & ~~1200$^{\dag}$ \\
     ~~~~10 & 12:58 & 16 & 3.2 & ~1* & + & 0 & 70--100 & ~~1100$^{\dag}$ \\
     ~~~~13 & 11:09 & 19 & 3.6 & 1~\, & + & 2 & \,Type I: 120~~~~~~~~ & ~700 \\
\multicolumn{7}{c}{} & Type II: 180--420+ \\
     ~~~~17 & 04:46 & 23 & 3.6 & 0~\, & + & 37 & 12--28 & ~450 \\
     ~~~~20 & 06:30 & 26 & 3.8 & 0~\, & + & 71 & 6--12 & ~220 \\
     ~~~~24 & 01:52 & 30 & 3.8 & 0~\, & + & 7 & 16--20 & ~220 \\
     ~~~~29 & 06:53 & 35 & 2.8 & 0~\, & + & 1 & 12 & ~230 \\
     ~~~~30 & 03:43 & 36 & 1.5 & 0~\, & + & 0 & 10--20 & ~230 \\
\\
1998 Jan~30 & 17:12 & ~3 & 6.5 & ~5* & + & 0 & 80--230 & 4000 \\
     ~~~~31 & 22:57 & ~4 & 3.5 & 2~\, & + & 0 & 180--220 & 2820 \\
     Feb~02 & 16:27 & ~6 & 6.1 & 2~\, & + & 0 & 220--240 & 2370 \\
     ~~~~04 & 19:48 & ~8 & 3.5 & 1~\, & + & 0 & 240 & 2100 \\
     ~~~~07 & 18:22 & 11 & 6.2 & 2~\, & + & 0 & 160--170 & 1030 \\
     ~~~~10 & 21:26 & 14 & 6.6 & 1~\, & + & 0 & 170 & 1040 \\
     ~~~~16 & 13:59 & 20 & 6.2 & 2~\, & + & 234 & \,Type I: 40--50 & ~240 \\
\multicolumn{7}{c}{} & Type II: 8--40~~ \\
     ~~~~19 & 11:53 & 23 & 8.1 & 1~\, & + & 91 & ~Type I: 50~~~~~ & ~250 \\
\multicolumn{7}{c}{} & Type II: 10--30 \\
\hline
\\
\multicolumn{9}{l}{\small * One of these bursts was observed during a slew.}\\
\multicolumn{9}{l}{\small $^{\dag}$ Count rates are estimated; no
offset pointing performed during these observations.}\\
\end{tabular}
\end{minipage}
\end{table*}
Timing and spectral analyses were conducted with data collected with
the Proportional Counter Array (PCA).  The PCA consists of five
identical xenon/methane proportional counters with a total effective
area of approximately 6500 cm$^{2}$; it is sensitive to X-rays in the
range 2--60 keV, and is capable of tagging relative event arrivals
down to 1 $\mu$s (Zhang et al. 1993).  Our observations used
individually described, event-encoded data with a time resolution of
122~$\mu$s and 64 energy channels.  The two standard data modes,
providing $\frac{1}{8}$~s timing data and 16~s/129 energy channel
data, were also available throughout.

The RB has been monitored almost continuously (except for $\sim$1
month per year when the source lies too close to the Sun) by the
\rxtez\ ASM since February 1996.  The ASM consists of three identical
Scanning Shadow Cameras (SSCs) mounted on a rotating assembly.  Each
SSC contains a position-sensitive proportional counter, which views
the sky through a coded mask. The ASM is sensitive to X-rays in the
range 2--10 keV.  ASM data is taken in a series of 90 second
``dwells'', with any randomly selected source being scanned typically
5--10 times per day (Levine et al. 1996).

During the May 1996 outburst, the RB was observed with the PCA on
three occasions near the end of the outburst (days 21--31 of the
outburst) (Lewin et al. 1996b).  On May 3 and May 7, 23 type~II bursts
of duration 8--17 seconds were observed.  Bursts occurred every
80--100 seconds on May 3, and every 300--600 seconds on May 7.  On May
13, no bursts were observed.

The RB was again observed with the PCA in November 1996, on days 8--19
of the outburst (Lewin et al. 1996c).  One type~I burst was observed
during each of the observations on November 6, 9 and 11.  The RB had
an average persistent emission (PE) level on those three occasions of
2270 cts s$^{-1}$, 1840 cts s$^{-1}$, and 1590 cts s$^{-1}$,
respectively.  No bursts were observed on November 10 and 17.  No PCA
observations were possible after November 17, due to the \rxtez\
Sun-angle constraint.

In June and July of 1997, the RB was observed at regular intervals
throughout an entire outburst for the first time (Guerriero et
al. 1997).  Type~I bursts were observed on days 2--17 of the outburst.
On July 13, two ``flat-topped'' type~II bursts of long duration (120
and $>$420 seconds, respectively) were observed.  Rapid type~II bursts
were then observed until day 36 of the outburst, when the outburst
ended (July 30).  The PE level declined steadily throughout the
outburst.

A second complete outburst was observed in January and February 1998
(Guerriero et al. 1998).  Observations with the PCA began on day 3 of
the outburst, and type~I bursts were seen exclusively through day 14.
On February 16, day 20 of the outburst, the RB was in a mode
characterized by many rapid type~II bursts, followed by a larger type
II burst.  This mode is identical to the mode in which the RB was
discovered (Lewin et al. 1976).  Rapid, regular type~II bursting
continued on February 19, the final observation of the outburst.

We have performed a spectral analysis to determine the conversion from
count rates to fluxes and to draw some rudimentary conclusions about
the X-ray emitting regions.  We emphasize that the model we present is
probably not uniquely indicated by the data.  No single-component
models provided acceptable fits to the data (minimum reduced
chi-squared values, for 51 degrees of freedom, are \chinu\ $\sim$ 3
for type~I bursts, 10 for type~II bursts, and 170 for the PE).  Two-
and three-component models that we considered incorporated blackbody,
disk blackbody, thermal bremsstrahlung, Comptonization, and power law
spectral components.  The best fitting models combined a power law
with two blackbody components, resulting in \chinu\ = 0.8--1.4.  A
simple two-component blackbody model, without a power law component,
also provided statistically acceptable fits to most of the type~I
bursts, but for many type~II bursts and the PE this model was not
acceptable (\chinu\ = 6.8 in some cases).  The addition of a power law
component improved the fits in these cases.  None of the other models
that we investigated provided statistically acceptable fits to the
data.

Best-fit temperatures of the two blackbody components were 1.1--1.5
keV and 0.25--0.40 keV, respectively.  Although the temperature of the
hotter component in the type~I bursts varied from burst to burst, it
cooled by $\sim$0.2 keV over the course of the type~I burst in nearly
every case (see Section~\ref{sec:I}).  The temperature of the cooler
component remained relatively constant during type~I bursts.  The
luminosity of the hotter component was $\sim$15 per cent that of the
cooler component (2.5--20 keV), in the bursts and in the PE.  When it
was necessary to include a power law component in the model, the
values for the photon index of the power law ranged from 1.9--4.0, and
luminosities ranged from 1--2 per cent of the cooler component
(2.5--20 keV).  Most of the type~II bursts and PE required a power law
component in the spectral model to obtain an acceptable fit to the
data.  The neutral hydrogen column density for our models was fixed at
2$\times$10$^{22}$ cm$^{-2}$, as determined from {\it EXOSAT\/}
observations (Tan et al. 1991); we did not find the low-energy
spectral response of the PCA sufficient, in these observations, to
constrain the column density independent of the other spectral
parameters.

One possible physical picture suggested by this model is of a system
with three X-ray emitting regions: the neutron star surface ($\sim$1
keV blackbody), the accretion disk ($\sim$0.3 keV blackbody), and a
Comptonizing cloud of hot electrons (responsible for the power law
component, when present).  The cooling of the hotter blackbody
component during type~I bursts is consistent with a cooling neutron
star surface (LVT93).  In contrast, the roughly constant temperatures
of the two blackbody components throughout the type~II bursts suggest
emitting regions that vary in size during the burst (normalizations of
both components vary with the X-ray flux).  We derive blackbody radii
for the hotter component during the type~I bursts of 9$\pm$2 to
14$\pm$2 $d_{8}$ km, depending on the burst, where $d_{8}$ is defined
as $D$/8~kpc and $D$ is the distance to the RB.  During type~I bursts
the cooler component has a blackbody emitting area of 0.7$\pm$0.3 to
2.3$\pm$1.0 $\times$10$^{6}$ $d_{8}^{2}$ km$^{2}$.

For the type~II bursts, we derive blackbody radii at the peak of the
bursts of 5.0$\pm$1.0 to 10$\pm$1.5 $d_{8}$ km, depending on the
burst, for the hotter component, and blackbody emitting areas of
0.5$\pm$0.3 to 2.1$\pm$0.9 $\times$10$^{6}$ $d_{8}^{2}$ km$^{2}$ for
the cooler component.  The normalizations of both components vary with
total X-ray flux during the bursts.

For the persistent emission spectra, we derive blackbody radii for the
hotter component of 10$\pm$2 to 13$\pm$2 $d_{8}$ km and blackbody
emitting areas of 0.9$\pm$0.4 to 2.5$\pm$0.8 $\times$10$^{6}$
$d_{8}^{2}$ km$^{2}$ for the cooler component.  The normalizations of
both components decrease over the course of the outburst as the total
X-ray flux decreases.

There are known problems with interpreting blackbody X-ray spectral
fits in such a literal fashion (LVT93).  One problem discussed in
LVT93 is that the observed X-ray color temperature and the effective
temperature (that is, the temperature if the source were a true
Planckian emitter) can differ by as much as a factor of $\sim$1.5.  In
the present case, this could lead to blackbody radii $1.5^{2} \sim 2$
times larger than the values we have quoted.

Bursts were classified as type~I or type~II, in part, by performing
spectral fits on the ``excess'' (PE-subtracted) burst counts.  This is
not a perfectly straightforward procedure (Van Paradijs \& Lewin
1985); however, while both type~I and type~II bursts can show some
spectral evolution during the burst, the spectral softening during the
decay is much more pronounced in a type~I burst (cf.\
Figure~\ref{fig:spec}).  The burst profile of a type~I burst, with a
sharp rise and a roughly exponential decay, is also substantially
different from that of a type II burst (except when the PE is near its
peak level, when burst profiles alone are not sufficient to
distinguish the two types).  Together with the spectral fits, then,
the burst profiles were used to classify bursts as either type~I or
type~II.  

The bright LMXB \slowb\ lies only 0.5$^\circ$ from the RB and is in
the field of view of the PCA when the RB is in the center of the field
of view.  To determine the contribution of this source, the satellite
pointing was offset by 0.5$^\circ$ away from \slowb\ for the last
third of each observation.  This procedure allowed us to estimate the
number of counts arriving from each source, but reduced our count rate
from the RB by about a factor of 2.5 for the offset phase of each
observation.  Additionally, an occasional type~I burst from \slowb\
was observed while the PCA was pointed directly at the RB.  These
bursts were easily distinguished from RB bursts by their peak flux
($\sim$15,000 PCA cts s$^{-1}$) and characteristic light curves, both
of which differ markedly from bursts emitted by the RB.  We also
detected the well-known 363 Hz oscillations (Strohmayer et al.\ 1996)
from several of the \slowb\ type~I bursts.

\section{Outburst Evolution}
\label{sec:evolution}

With the \rxtez\ ASM we can detect the onset of a RB outburst to
within less than one day.  The four outbursts observed with \rxtez\
began on 13 April 1996, 30 October 1996, 25 June 1997, and 28 January
1998.  The intervals between the start of these outbursts are 200,
238, and 217 days, respectively.  Using these three values, we find
the current average recurrence time for RB outbursts to be 218 days.

% This is now Figure 2
\begin{figure}
\begin{center}
~\psfig{figure=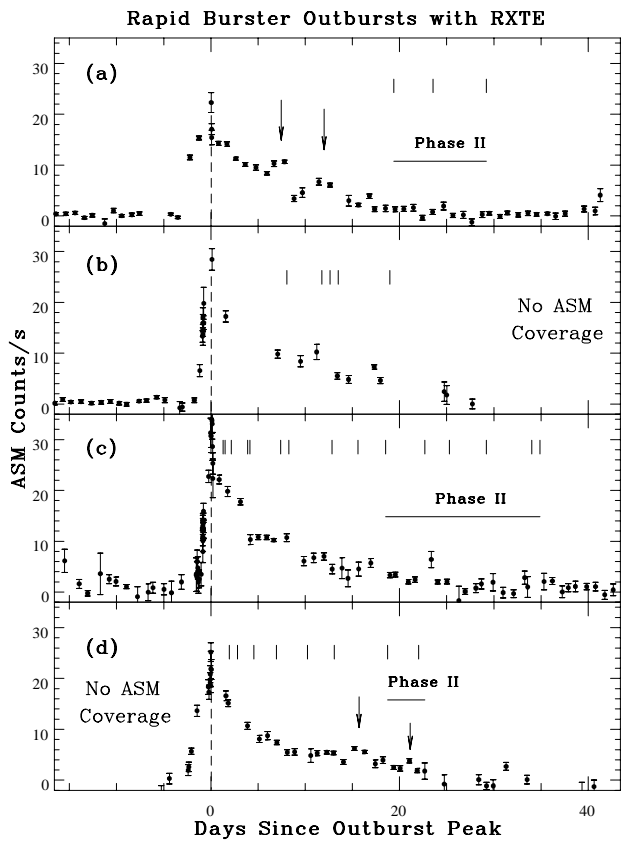,bbllx=43pt,bblly=35pt,bburx=565pt,bbury=760pt,width=3.2in}~
\caption{One-day averages of ASM count rates for outbursts of the
Rapid Burster (90 second data points are also plotted during the
outburst rise).  The peaks of the outbursts have been aligned.  The
data are from (a) 29 March 1996 -- 27 May 1996 (outburst peak: 14 Apr
96).  (b) 15 October 1996 -- 18 November 1996 (outburst peak: 31 Oct
96).  (c) 10 June 1997 -- 8 August 1997 (outburst peak: 26 Jun 97).
(d) 25 January 1998 -- 13 March 1998 (outburst peak: 30 Jan 98).
Short vertical lines indicate days when PCA observations were made.
Horizontal lines indicate periods in which the RB was observed to be
in Phase II (type~II burst dominated); during other periods the RB was
in Phase I (type~I burst dominated).  Arrows point to the fitted
centre of a Gaussian curve which may indicate the presence of a
reflare (see text).  Prior to the start of an outburst, the RB flux
detected with the ASM is consistent with zero.}
\label{fig:asm}
\end{center} 
\end{figure}

\begin{table*}
\begin{minipage}{6.5in}
  \centering
\caption{Parameters describing the best-fit curve of each Rapid
  Burster outburst.  The outburst rise is linear, while during
  outburst decay the ASM flux $\propto$ $e^{-t/\tau}$.  The
  times of the additional peaks (modeled as Gaussians) are given in
  days since the primary peak.  We do not identify any additional peaks
  during the November 1996 or June/July 1997 outbursts.}
\label{tbl:fits}
\begin{tabular}{cccccccc}  
\\
\multicolumn{5}{c}{} & \multicolumn{3}{c}{Additional Peaks} \\
\cline{6-8}
\multicolumn{1}{c}{}  &  Start Time  &  Rise Time  & 
    Peak Flux  &  $\tau$  &  Centre  &  Amplitude  &  Width \\
Outburst & ($t_{o}$, MJD) & ($t_{p}-t_{o}$, days) & 
    ($P$, ASM cts s$^{-1}$) & (days) & ($t_{n}-t_{p}$, days) & 
    ($A_{n}$, ASM cts s$^{-1}$) & ($\sigma_{n}$, days) \\
\hline
May 1996 & 50186.2$\pm$2.0 & 2.4$\pm$0.04 & 16$\pm$2 & 7.9$\pm$0.5 & 
    ~8$\pm$0.5 & 3.0$\pm$0.2 & 0.5$\pm$0.2\\
\qq & \multicolumn{4}{c}{} & 13$\pm$0.5 & 2.8$\pm$0.2 & 0.5$\pm$0.2\\
Nov 1996 & 50386.0$\pm$1.2 & 1.9$\pm$0.03 & 26$\pm$2 & 8.6$\pm$0.7 & 
    -- & -- & -- \\
Jun 1997 & 50624.1$\pm$1.5 & 1.7$\pm$0.16 & 26$\pm$2 & 8.5$\pm$0.8 & 
    -- & -- & -- \\
Jan 1998 & 50841.4$\pm$1.4 & 3.1$\pm$0.06 & 21$\pm$2 & 6.0$\pm$0.7 & 
    17$\pm$1.0 & 5.7$\pm$0.5 & 3.3$\pm$0.7 \\
\qq & \multicolumn{4}{c}{} & 23$\pm$0.5 & 1.6$\pm$0.6 & 0.4$\pm$0.3\\
\hline
\end{tabular}
\end{minipage}
\end{table*}

The four most recent outbursts have all followed a nearly identical
evolution in the ASM (Figure~\ref{fig:asm}).  The time-averaged X-ray
flux (PE and bursts) observed with the ASM rises linearly at the
beginning of an outburst and peaks within $\sim$1--3 days.  The
average X-ray flux then declines exponentially over the next $\sim$35
days.  During the May 1996 and January/February 1998 outbursts, there
are indications of reflares.  In both cases, in addition to the main
peak, two additional peaks are preferred, as determined by F-tests
(F-test probabilities with two additional peaks are 99.8 per cent and
99.3 per cent, respectively).  There is no evidence for additional
peaks during the November 1996 or June/July 1997 outbursts.  These
reflares are reminiscent of the behavior of soft X-ray transients
(Augusteijn, Kuulkers, \& Shaham 1993; for a review of soft X-ray
transients, see Tanaka \& Lewin 1995).  During a one-day period
between days 5 and 6 of the June/July 1997 outburst (June 29 and June
30, 1997), there is a sudden decrease in the average flux from the
source (see Figure~\ref{fig:asm}c).

An outburst can be parameterized by:
\[ f(t) = \left\{ \begin{array}{ll}
	\frac{P}{t_{p} - t_{o}} ~(t - t_{o}), & \mbox{$t_{o} < t \leq
	t_{p}$} \\ 
	P e^{-(t - t_{p})/\tau} + \sum_{n} A_{n} \exp [-
	\frac{(t-t_{n})^{2}}{2 \sigma_{n}^{2}}], & \mbox{$t > t_{p}$}
	\end{array} \right. \]

Here, $P$ is the peak flux (ASM cts s$^{-1}$), $t_{o}$ is the time of
the outburst start (days), $t_{p}$ is the time of outburst peak
(days), and $n$ parameterizes the additional peaks, with $A_{n}$ the
amplitude of the $n$th additional peak (in ASM cts s$^{-1}$), $t_{n}$
its peak time (days), and $\sigma_{n}$ its Gaussian width.  The fitted
parameters for each outburst are summarized in Table~\ref{tbl:fits}.

Our PCA observations of the RB occurred intermittently during four
outbursts.  The source was observed at varying intervals during each
outburst.  All four sets of observations, however, are consistent with
one global evolutionary pattern for a RB outburst.  A RB outburst can
be divided into two phases: Phase I is dominated by type~I bursts with
a strong PE, lasting for 15--20 days.  The PE declines steadily during
this phase.  Phase II consists of several different modes of type~II
bursting, and lasts until the end of the outburst.
Table~\ref{tbl:evo} summarizes the evolution of the RB outbursts.

\subsection{Phase I}
\label{sec:I}

Within the first 1--3 days of a RB outburst, the persistent emission
level rises quickly to its peak level of $\sim$5000 PCA cts s$^{-1}$
(1 count s$^{-1}$ $\simeq$ 3$\times$10$^{-12}$ erg cm$^{-2}$ s$^{-1}$,
2--20 keV).  Although there have been no PCA observations in this
initial rising phase, there have been a few observations near the peak
of an outburst. The PE level does not remain constant for any
appreciable amount of time; it declines steadily during Phase I from
its peak level down to $\sim$1000 cts s$^{-1}$ by the end of the
phase.

\begin{table*}
\begin{minipage}{5.75in}
  \centering
\caption{The evolution of Rapid Burster outbursts observed by \rxtez\
  (1 PCA count s$^{-1}$ $\simeq$ 3$\times$10$^{-12}$
  erg cm$^{-2}$ s$^{-1}$; 2--20 keV).}
\label{tbl:evo}
  \begin{tabular}{cccl}
\\
\multicolumn{1}{c}{Days of} & & PE Level\\ 
\multicolumn{1}{c}{Outburst} &
\multicolumn{1}{c}{Phase} & 
\multicolumn{1}{c}{(PCA cts s$^{-1}$)} &
\multicolumn{1}{l}{~~Burst Behavior} \\ 
\hline
~1--17 & I & 5000 $\rightarrow$ 1000 & Type I bursts \\
\multicolumn{3}{c}{} & ~~200--250 s duration \\
\multicolumn{3}{c}{} & ~~1500--3000 s between bursts \\
\multicolumn{3}{c}{} & Possible type II bursts\\
\hline
18--19 & II & 1000 $\rightarrow$ 700 & Type II bursts \\
\multicolumn{1}{c}{} & Mode 0 & \multicolumn{1}{c}{} & 
~~long, flat-topped bursts \\
\multicolumn{3}{c}{} & ~~100--700 s duration\\
\multicolumn{3}{c}{} & Occasional type I bursts\\
\hline
20--21 & II & 200--500 & Type II Bursts \\
\multicolumn{1}{c}{} & Mode I & \multicolumn{1}{c}{} & 
~~8--40 rapid bursts (8--12 s duration), \\
\multicolumn{3}{c}{} & ~~~followed by a large burst (20--30 s duration)\\
\multicolumn{3}{c}{} & ~~``ringing'' during burst decay \\
\multicolumn{3}{c}{} & Small type I bursts may follow\\
\multicolumn{3}{c}{} & ~~large type II bursts\\
\hline
22--35 & II & 200--500 & Type II Bursts \\
\multicolumn{1}{c}{} & Mode II & \multicolumn{1}{c}{} & 
~~burst intervals can be very regular\\
\multicolumn{3}{c}{} & ~~40--100 s between bursts\\
\multicolumn{3}{c}{} & ~~burst durations can vary from 5--25 s\\
\multicolumn{3}{c}{} & ~~``ringing'' during burst decay \\
\multicolumn{3}{c}{} & Occasional small type I bursts\\
\hline
\end{tabular}
\end{minipage}
\end{table*}

\begin{figure}
\begin{center}
~\psfig{figure=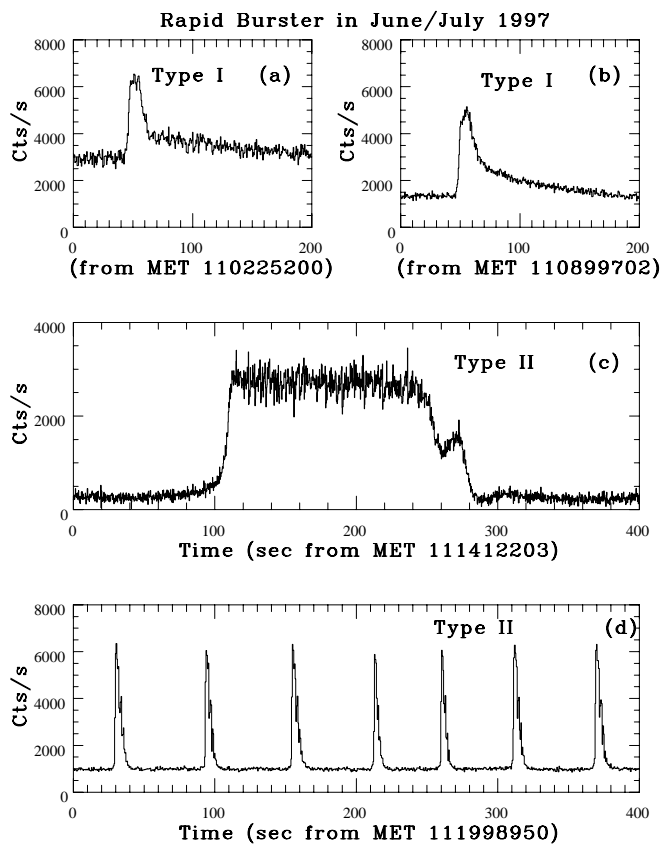,bbllx=43pt,bblly=35pt,bburx=565pt,bbury=760pt,width=3.2in}~
\caption{The Rapid Burster during the outburst of June/July 1997.  (a)
\& (b) Type~I bursts during Phase I (days 5 \& 13 of the outburst).
Notice the long tails during the burst decay.  (c) Long type~II burst
during Mode 0 of Phase II (day 19 of the outburst; notice the dip in
the PE following the burst).  (d) Type~II bursts during Mode II of
Phase II (day 23 of the outburst).  Contributions from 4U 1728-34 have
been removed.  Figure~(b) has been corrected for aspect; this was
necessary as that burst occurred during an offset pointing (see
text).}
\label{fig:jun}
\end{center} 
\end{figure}

We observed Phase I to last for 15--20 days.  The bursts during this
phase that could be positively identified were exclusively type~I
bursts. The RB type~I bursts have a profile similar to type~I bursts
observed from some other LMXBs (Figures~\ref{fig:jun}a \&
\ref{fig:jun}b).  Typical burst durations were 200--250 seconds, with
$\sim$1500--3000 seconds between bursts.  We observed peak excess
count rates for these bursts to be typically 2000--5000 cts s$^{-1}$.
The rise times for the bursts decrease as Phase I progresses.
Initially, the rise times were 10--20 seconds; however, by day 3 of
the outburst the rise times had decreased to 3--5 seconds.

% This is now Figure 3

\begin{figure}
\begin{center}
~\psfig{figure=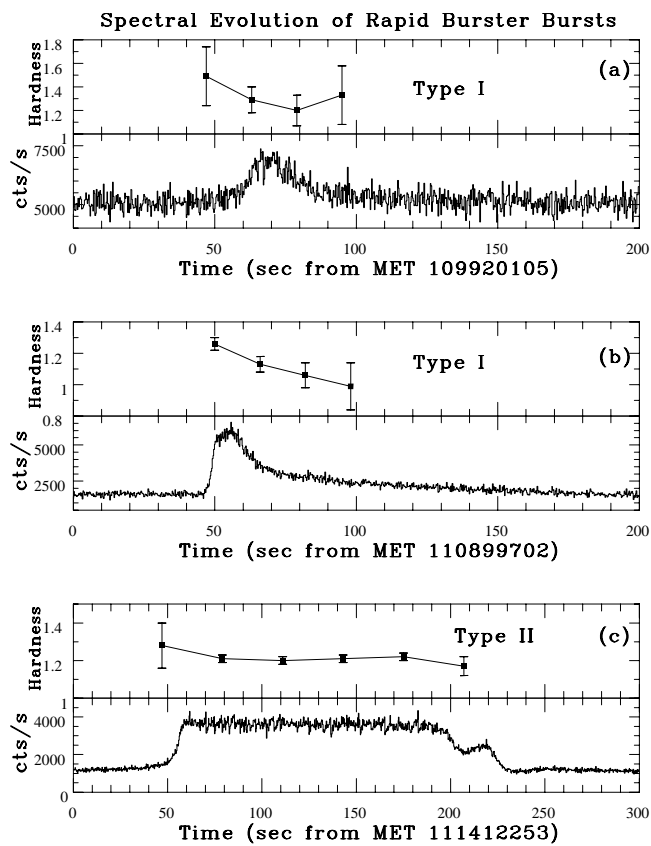,bbllx=43pt,bblly=35pt,bburx=565pt,bbury=760pt,width=3.2in}~
\caption{Spectral evolution of Rapid Burster bursts.  Hardness is a
measure of the blackbody temperature of the excess burst counts above
persistent emission.  (a) A type~I burst that occurred on June 26,
1997 (day 2 of the outburst).  When the persistent emission is at such
a high level, type~I bursts do not always have their characteristic
profile.  (b) A type~I burst that occurred on July 7, 1997 (day 13 of
the outburst).  Note the drop in the persistent emission level since
the burst shown in (a); also note the long tail during the burst
decay.  (c) A long type~II burst that occurred on July 13, 1997 (day
19 of the outburst).  Note the lack of spectral evolution and the
shape of the burst, making it unquestionably a type~II burst (see Tan
et al. 1991).}
\label{fig:spec}
\end{center} 
\end{figure}

\begin{figure}
\begin{center}
~\psfig{figure=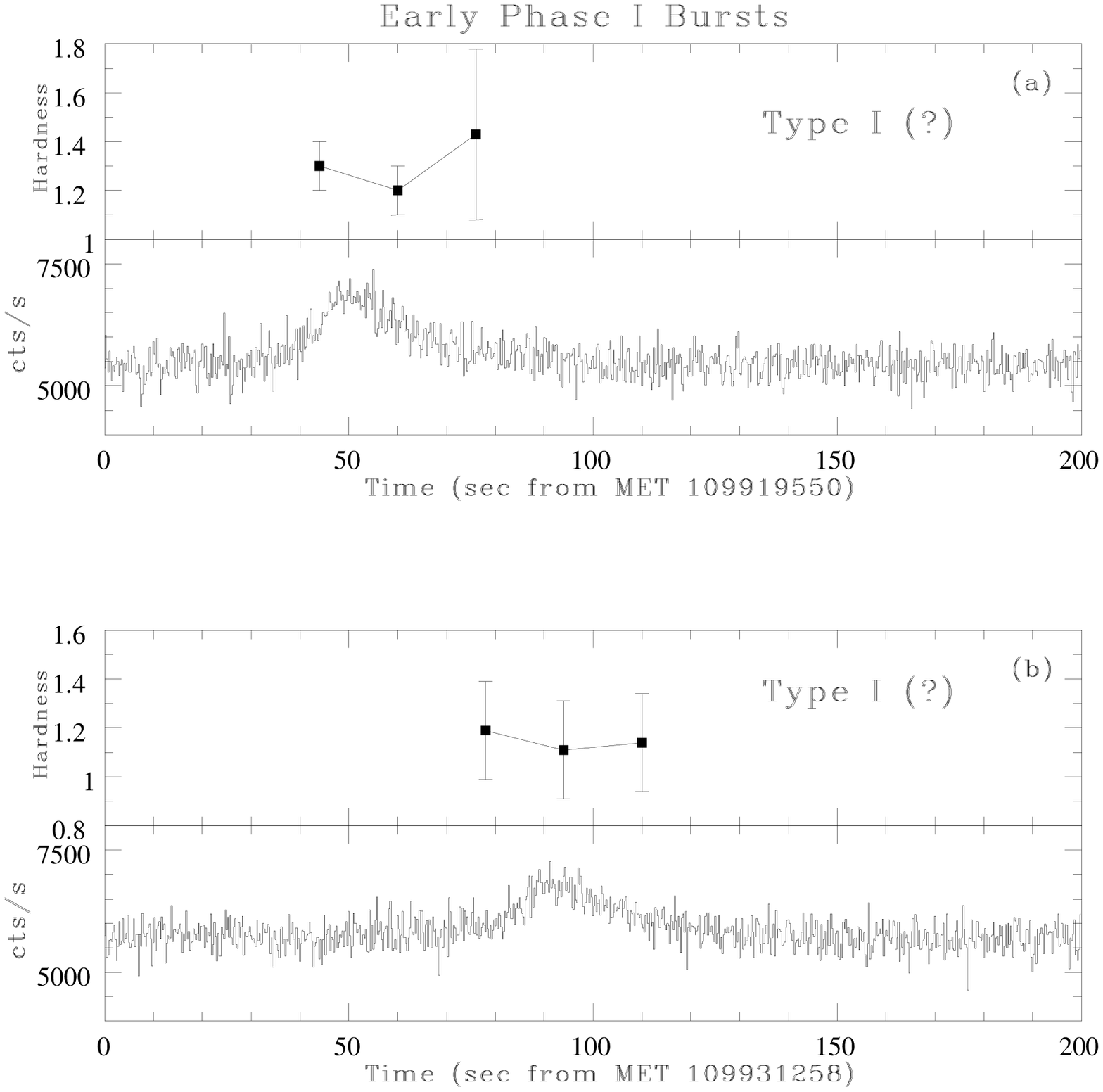,bbllx=43pt,bblly=150pt,bburx=565pt,bbury=700pt,width=3.2in}~
\caption{Two bursts from the early stages of Phase I (26 June 1997,
day 2 of the outburst).  Hardness is a measure of the blackbody
temperature of the excess burst counts above persistent emission.  The
persistent emission level was at its peak during this portion of the
outburst.  The burst profiles, which are not easily distinguishable as
either type~I or type~II burst profiles, and the poorly constrained
spectral evolution make these bursts difficult to classify as type~I
or type~II bursts.  Note the slow rise of both bursts.}
\label{fig:ambig}
\end{center} 
\end{figure}

The spectra of 33 out of 35 bursts observed during this phase soften
during the burst decay, indicating that they are type~I bursts
(Figures~\ref{fig:spec}a \& \ref{fig:spec}b).  Two bursts did not show
detectable spectral softening, and are therefore difficult to classify
as either type~I or type~II (Figure~\ref{fig:ambig}).  These bursts
occurred on day 2 of the June/July 1997 outburst (June 26, 1997), when
the PE was at its highest level observed with the PCA ($\sim$5000 cts
s$^{-1}$); moreover, both bursts had peak excess count rates of $<$
1500 PCA cts s$^{-1}$.  It is possible that these were type~II bursts.
However, given that these two bursts resemble the other bursts
observed during the early stages of Phase I, we feel that they are
most likely type~I bursts.

% This is now Figure 4

\subsection{Phase II}

We consider Phase II, which is dominated by type~II bursts, to begin
with the first type~II burst.  Phase II can be generally divided into
three sub-phases, each having a different pattern of type~II bursts.
Although the start of type~II bursting indicates the beginning of
Phase II, type~I bursting does not cease.  Type~I bursts identical to
those from the first phase can be seen during the initial stages of
Phase II.  In addition, short type~I bursts can be observed throughout
Phase II.  Although we will describe these sub-phases in the sequence
that we observed them to occur, it is important to note that
historically the RB has not always followed the same sequence (see
Section~\ref{sec:historical}).

On day 18 of the June/July outburst, the RB produced two long,
flat-topped type~II bursts.  These bursts had durations of 180 seconds
and $>$420 seconds (this burst was observed in progress as the
spacecraft slewed onto the source).  There was very little spectral
evolution during the bursts (Figure~\ref{fig:jun}c).  Dips were
present in the PE following both of the bursts, although there was no
dip prior to the only complete burst we observed in this phase.  In
the last 30 seconds of the shorter burst, the familiar ``ringing'' of
RB type~II bursts was evident (Figure~\ref{fig:jun}c).  This
``ringing'' during burst decay is only clearly evident if the bursts
are less than $\sim$200 seconds in duration (Tan et al. 1991).  This
pattern of flat-topped type~II bursts was first observed with \hak\ in
August 1979, and was called ``Phase II'' by Kunieda et al.  (1984a).
This should not be confused with our use of Phase I and Phase II,
which describe phases of type~I burst-dominated and type~II
burst-dominated emission, respectively.  We propose instead to
designate this mode (type~II bursts $>$100 seconds) as ``Mode 0'', to
indicate the fact that it appears to be the initial mode of Phase II.
On July 13, 1997, a type~I burst was observed while the RB was in Mode
0.

\begin{figure}
\begin{center}
~\psfig{figure=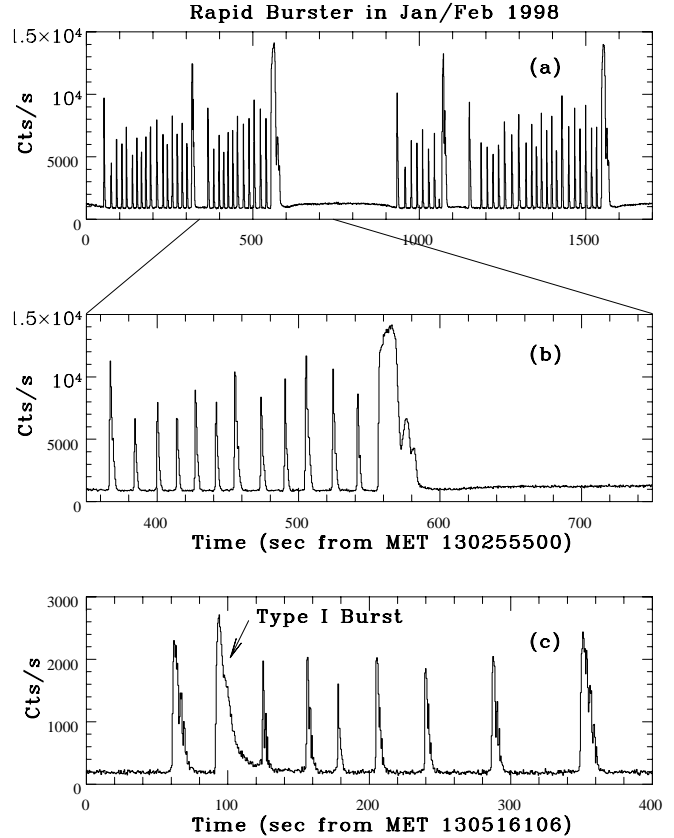,bbllx=43pt,bblly=35pt,bburx=565pt,bbury=760pt,width=3.2in}~
\caption{Rapid Burster bursts from the January/February outburst of
1998.  (a) Phase II, Mode I (day 20 of the outburst).  Note the
enhanced emission following the large type~II bursts.  (b) A 400
second segment of Figure~\ref{fig:jan}(a).  (c) Bursts during Phase
II, Mode II (day 23 of the outburst).  Note the type~II burst that
occurred during the decay of a type~I burst (at time $\sim$100
s). Also note the ``ringing'' that can be seen during type~II burst
decay in all three figures.}
\label{fig:jan}
\end{center} 
\end{figure}

The next burst pattern is characterized by several (8-40) short type
II bursts, followed by a larger type~II burst (Figures~\ref{fig:jan}a
\& \ref{fig:jan}b).  Following Marshall et al. (1979), we refer to
this as ``Mode I''.  We observed the RB emission in this pattern on
February 16, 1998, which was day 20 of that outburst.  There was a
long delay following each of the large type~II bursts before the
occurrence of the next type~II burst.  The shorter bursts had
durations of 8--12 seconds, with $\sim$10 seconds between bursts,
while the longer bursts lasted for 20--30 seconds.  The pause
following a large type~II burst varied from 50 to 350 seconds.  There
was also a period of enhanced X-ray emission following the large type
II bursts.  All of the type~II bursts observed during this phase
exhibited the timescale-invariant profile.  Type~I bursts can also
occur in this phase, usually during the PE period immediately
following a large type~II burst.  This mode is the one observed when
the RB was discovered by Lewin et al. (1976).

% Now figure 5

The final pattern is made up of many short type~II bursts in a nearly
regular pattern (Figures~\ref{fig:jun}d \& \ref{fig:jan}c), which was
designated ``Mode II'' by Marshall et al. (1979).  We observed this
mode as the final phase before the end of each outburst seen with
\rxtez\ (we did not observe the end of the outburst of the RB in
November, 1996, since it was too close to the Sun to be observed).
The burst profiles are all timescale-invariant, and exhibit
``ringing'' during burst decay (Tawara et al. 1985; Tan et al. 1991).
Burst durations are 5--25 seconds, with bursts occurring every 40--100
seconds.  We did observe type~I bursts during this phase, some of
which occur simultaneously with type~II bursts
(Figure~\ref{fig:jan}c).  At the latest point in an outburst in which
we observed Mode II (July 24, 1997; day 30 of the outburst), the burst
separation had increased to $\sim$500 seconds.

\begin{figure}
\begin{center}
~\psfig{figure=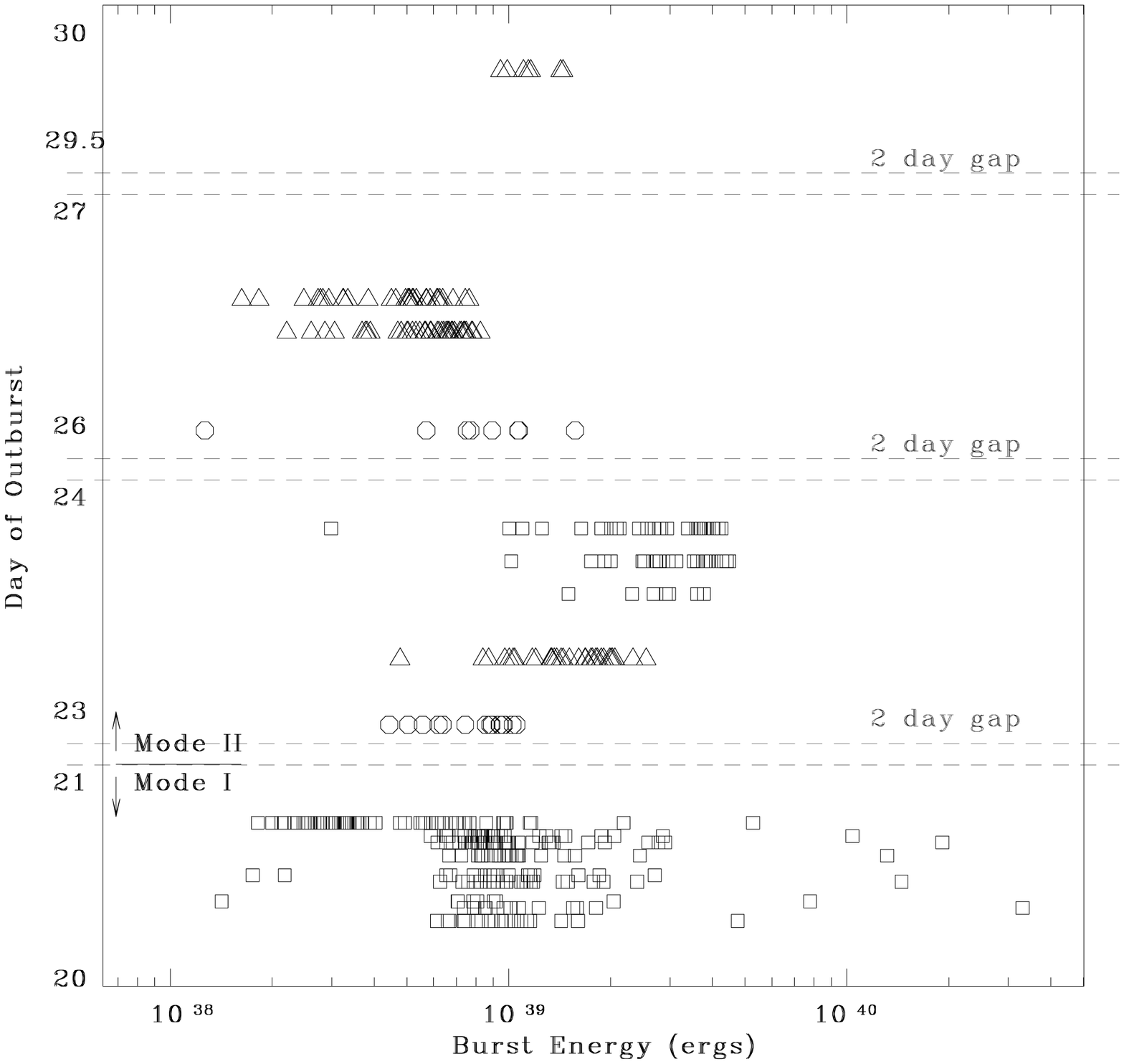,bbllx=43pt,bblly=150pt,bburx=565pt,bbury=700pt,width=3.2in}~
\caption{The total energy in all type~II bursts observed with {\textit
RXTE}.  The total energy in each burst (2--20 keV, assuming isotropic
emission and a distance of 8~kpc) is plotted against the approximate
time of its occurrence.  Circles indicate type~II bursts of the May
1996 outburst; triangles indicate type~II bursts of the June/July 1997
outburst; squares indicate type~II bursts of the January/February 1998
outburst.  There were no type~II bursts observed during the November
1996 outburst.  The transition from Mode I to Mode II occurs at
approximately day 22 of these outbursts.  The results shown in this
figure show a striking similarity to those presented by Marshall et
al. (1979) in a similar figure.  The reader is encouraged to consult
the original figure by Marshall et al. (1979).}
\label{fig:herm}
\end{center} 
\end{figure}

Mode I and Mode II are distinguishable in Figure~\ref{fig:herm}.  The
burst energy $E$ has been plotted for each burst at the approximate
day of the outburst in which the burst occurred.  There is a bi-modal
distribution of burst energies during Mode I, while Mode II exhibits a
single-peaked distribution.
\vspace{.3in}

% Figure 7

\subsection{$E$-$\Delta t$ Relation}

\begin{figure}
\begin{center}
~\psfig{figure=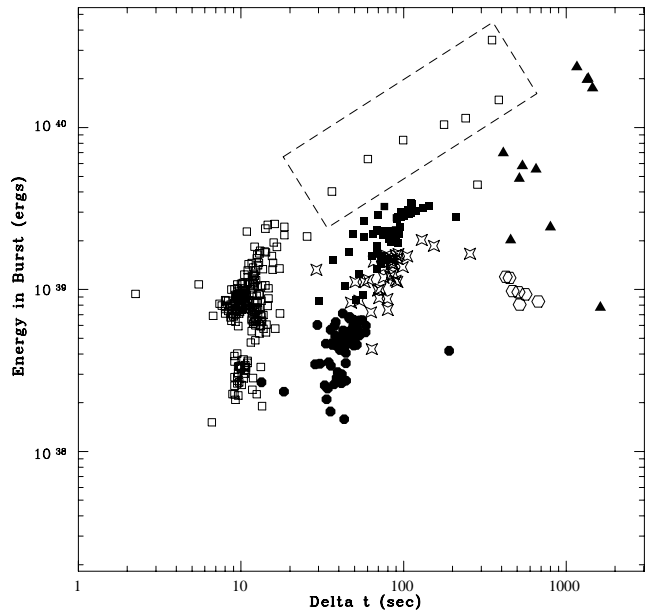,bbllx=43pt,bblly=150pt,bburx=565pt,bbury=700pt,width=3.2in}~
\caption{$E$-$\Delta t$ Relation.  We have assumed isotropic emission
and a source distance of 8 kpc.  Open squares are type~II bursts of
Mode I observed during February 1998 (day 20 of the outburst).  Those
in the boxed region are the large type~II bursts of the same mode that
were followed by a segment of enhanced PE.  Filled squares are type~II
bursts of Mode II from February 19, 1998 (day 23 of the outburst).
Open stars are type~II bursts from July 17, 1997 (also day 23 of the
outburst; Mode II).  Filled circles are type~II bursts of Mode II that
occurred on July 20, 1997 (day 26 of the outburst).  Open hexagons are
also type~II bursts of Mode II; they occurred on day 29 of the
outburst.  Filled triangles are type~I bursts from Phase I.  Note that
only bursts occurring on a given day may exhibit a single $E$-$\Delta
t$ relation, since the time-averaged X-ray luminosity
($\frac{E}{\Delta t}$) decreases during an outburst.}
\label{fig:Et}
\end{center} 
\end{figure}

Since its discovery, the RB has been known to have a proportional
relationship between the fluence of a type~II burst and the waiting
time to the {\it next} type~II burst (Lewin et al. 1976).  In this
sense, the RB behaves like a relaxation oscillator.  We have
calculated the total energy of 398 type~II bursts that occurred during
the June 1997 and January 1998 outbursts, assuming isotropic emission
and a source distance of 8 kpc.  These are plotted against the waiting
time to the next burst in Figure~\ref{fig:Et}.  In addition, eleven
type~I bursts from Phase I are included in the figure.

% Figure 8

In general, the relationship between the energy in a burst and the
waiting time to the next burst can be described by a power law of the
form $E$=$\beta (\frac{\Delta t}{100 s})^{\alpha}$ (Kunieda et
al. 1984a).  Here, we have defined $\beta$ to be the energy in a type
II burst that ``generates'' a 100 second waiting time to the next
burst.  In rare cases, this $E$-$\Delta t$ relation can be
approximately linear ($\alpha$=1).  One relation does not hold for all
bursts in an outburst, because the time-averaged type~II X-ray burst
luminosity ($\ell = \frac{E}{\Delta t}$) decreases during an outburst
(Figure~\ref{fig:asm}).  During our 2--4 ksec observations on any
given day, however, $\alpha$ and $\beta$ remained relatively constant.
The derived values of $\ell$ show a progressive decrease as the
outburst evolves.  Values for $\alpha$, $\beta$, and $\ell$ are
summarized in Table~\ref{tbl:param}.

\begin{table}
  \centering
\caption{$E$-$\Delta t$ relation parameters for type~II bursts, where
$\alpha$ is the power law index, $\beta$ is the energy in a type~II
burst that ``generates'' a 100 second waiting time to the next burst,
and $\ell$ is the time-averaged type~II burst luminosity (see also
Figure~\ref{fig:Et}).}
\label{tbl:param}
  \begin{tabular}{ccccc}
\multicolumn{1}{c}{} & Day of & & $\beta$ & $\ell$ \\
Date & Outburst & $\alpha$ & {\scriptsize (10$^{39}$ ergs)} & 
{\scriptsize (10$^{37}$ ergs s$^{-1}$)}\\
\hline
16 Feb & ~20$^{\dag}$ & 0.52$\pm$0.21 & 8.5$\pm$0.5 & 8.0$\pm$0.2 \\
19 Feb & 23 & 0.94$\pm$0.12 & 3.4$\pm$0.2 & 3.1$\pm$0.1 \\
17 Jul & 23 & 0.43$\pm$0.13 & 1.6$\pm$0.3 & 1.7$\pm$0.1 \\
20 Jul & 26 & 0.48$\pm$0.13 & 0.53$\pm$0.02 & 1.2$\pm$0.1 \\
\hline
\\
\multicolumn{5}{c}{\scriptsize $^{\dag}$Only includes the large
  bursts of Mode I (boxed region of Fig.~7).}
\end{tabular}
\end{table}

The short type~II bursts that occurred during Mode I appear to have a
nearly constant burst interval, $\Delta t$, over a large range of
fluences.  We do not have enough data points to define a meaningful
relation for the bursts that occur at the very end of an outburst
(open hexagons of Figure~\ref{fig:Et}).  Finally, the $E$-$\Delta t$
relation is not relevant for type~I bursts (filled triangles of
Figure~\ref{fig:Et}).

\section{Historical Perspective}
\label{sec:historical}

The RB has been observed intermittently since its discovery in 1976
(Figure~\ref{fig:cov}).  Various satellites have observed the RB at a
variety of wavelengths.  X-ray observations of the RB are summarized
in Table~\ref{tbl:sat}.

\begin{table}
  \centering
\caption{X-ray observations of Rapid Burster outbursts from
  1971--1998.  Phase(s) indicates the burst phase(s) observed with
  each satellite.  Mode(s) indicates the mode(s) observed when the RB
  was in Phase II.}
\label{tbl:sat}
  \begin{tabular}{cccc}
Obs Date & Satellite & Phase(s) & Mode(s) \\
\hline
Mar 71 & \uhur\ & \multicolumn{2}{c}{undetermined}\\
May 72 & \uhur\ & \multicolumn{2}{c}{undetermined}\\
Apr 73 & \cop\ & \multicolumn{2}{c}{undetermined}\\
Feb 75 & \cop\ & \multicolumn{2}{c}{undetermined}\\
Mar 76 & \sas\ & II & I\\
 \qq   & \ariel\ & II & II $\rightarrow$ I \\
Apr 77 & \ariel\ & II & II $\rightarrow$ I $\rightarrow$ II \\
Sep 77 & \sas\ & II & II \\
Mar 78 & \heao\ & II & I \\
 \qq   & \sas\ & II & II  \\
Oct 78 & \sas\ & II & II \\
Mar 79 & \sas\ & II & 0 $\rightarrow$ II \\
 \qq   & \ein\ & II & II \\
Aug 79 & \hak\ & II & 0 $\rightarrow$ II \\
Apr 83 & {\it ASTRON} & I &  \\
Aug 83 & \tenma\ & I & \\
 \qq   & {\it ASTRON} & I & \\
 \qq   & \exo\ & I $\rightarrow$ II & II \\
 \qq   & \hak\ & II & 0 $\rightarrow$ II \\
Jul 84 & \tenma\ & II & 0 $\rightarrow$ I $\rightarrow$ II \\
 \qq   & \exo\ & II & II \\
Aug 85 & \exo\ & II & 0 $\rightarrow$ II \\
Aug 88 & \ging\ & II & I $\rightarrow$ II \\
Jan 89 & \ging\ & ? \\
May 96 & \rxte\ & II & II \\
Nov 96 & \rxte\ & I & \\
Jun 97 & \rxte\ & I $\rightarrow$ II & 0 $\rightarrow$ II\\
Jan 98 & \rxte\ & I $\rightarrow$ II & I $\rightarrow$ II\\
\hline
\end{tabular}
\end{table}

\begin{figure}
\begin{center}
~\psfig{figure=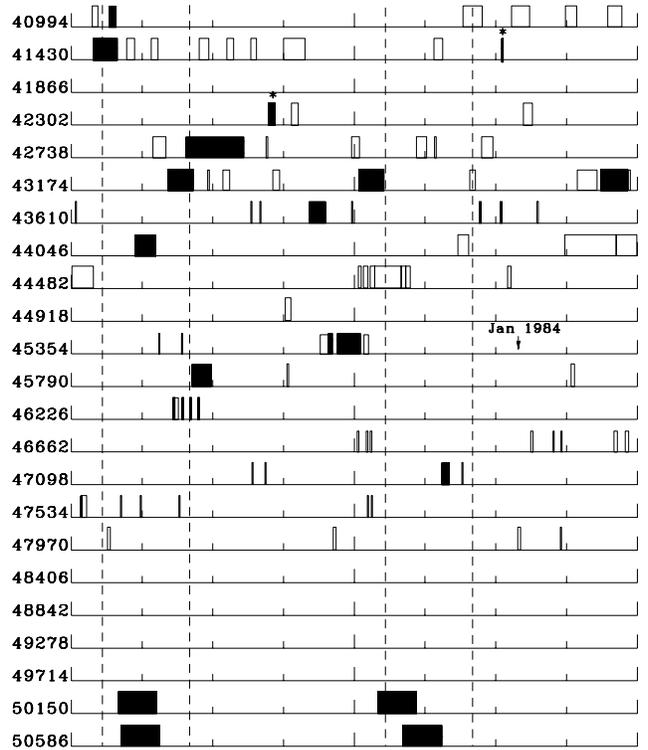,bbllx=43pt,bblly=35pt,bburx=565pt,bbury=760pt,width=3.2in}~
\caption{Observations of the Rapid Burster (1971-1998), folded with a
218 day period (2 cycles per line).  Modified Julian Dates (MJD) are
listed along the ordinate.  Filled boxes indicate observations that
occurred when the RB was active.  Open boxes indicate observations in
which the RB was inactive.  Single lines indicate observations of less
than one day in which the RB was inactive.  The observations marked
with ``*'' indicate periods in which the source may have been active,
but this is uncertain (White et al., 1978).  Dashed lines indicate 65
day windows.  MJD 40994 corresponds to February 12, 1971.}
\label{fig:fold}
\end{center} 
\end{figure}

Using the 218 day average period determined from the ASM data, we
folded the record of RB observations with this value
(Figure~\ref{fig:fold}).  The few outbursts that were observed from
1984--1989 seem to fit the pattern reasonably well, while those
outbursts prior to 1984 do not fit.  In fact, a good fit for the
outbursts from 1976--1983 is obtained with a 181 day period.  Other
periods, from 100--300 days, were also used to fold the data, but none
produced better agreement with the observed outbursts than the 181 and
218 day periods.  Although years have passed without any positive
detections of the RB in outburst, Figure~\ref{fig:fold} indicates that
very few of the observations during those years fell in the estimated
outburst windows.  It therefore seems likely that the RB goes into
outburst at semi-regular intervals of 6--8 months.  Continued
monitoring with the \rxtez\ ASM will be essential to confirm and
characterize this behavior.

% Figure 9

RB outbursts observed with other satellites have generally followed
the evolution that we have described, with some exceptions.  In all
observations in which the RB was in Phase I (Aug 83, Jun 97, Jan 98),
Phase II was observed several days later (Kunieda et al. 1984b).
There were two occasions, however, when Phase II may not have been
preceded by Phase I.  From March 8--15, 1978, \sas\ observations
indicated that the RB was not in outburst.  However, on March 18,
1978, both \sas\ and \heao\ observed the RB in Phase II (Mode I)
(Jernigan et al 1978).  \hak\ observations from July 31-August 7,
1979, detected no bursts from the RB.  On August 8, 1979, rapid
repetitive bursts were observed from the RB, indicating that it was in
Phase II (Mode II) (Kunieda et al. 1984a).  If Phase I occurred in
these cases, it must have been extremely short-lived.  In all other
observations of the RB in outburst, there is a large enough gap (two
weeks) between the last pre-outburst observation and the observation
of Phase II for Phase I to have occurred ``normally''.  It thus seems
that there is a strong indication that the Phase I to Phase II pattern
is characteristic of the RB outbursts.

In all RB observations, Phase II has been observed as the final phase
of the outburst.  There are indications that the RB follows the
progression within Phase II that we have described; that is `` Mode 0
$\rightarrow$ Mode I $\rightarrow$ Mode II''.  In July 1984, \tenma\
and then \exo\ observed the RB transition from Mode 0 to Mode I to
Mode II (Kawai et al. 1990; Lubin et al. 1991).  \heao\ and \sas\, in
March 1978, and \ging\, in August 1988, observed the transition from
Mode I to Mode II (Jernigan et al. 1978; Hoffman et al. 1978b; Dotani
et al. 1990).  Many satellites have observed Mode 0, and later Mode
II, without seeing Mode I in between.  These include \sas\ in March
1979 (Basinska et al. 1980), \hak\ in August 1979 and in August 1983
(Inoue et al. 1980; Kunieda et al. 1984a; Kunieda et al. 1984b), and
\exo\ in August 1985 (Stella et al. 1988a; Lubin et al. 1992b).
However, Mode I might have occurred between these observations, since
all of the observations were intermittent. All of these observations
saw Phase II, Mode II, as the final Mode before the end of the
outburst.

Dips in the persistent emission prior to and just following a type~II
burst are common when the RB is in Mode 0, as is enhanced PE following
the large type~II bursts of Mode I.  This enhanced PE during Mode I
can be clearly seen in the first \sas\ observation of the RB (Lewin et
al. 1976), in the \heao\ observation from March 1978 (Hoffman et
al. 1978b), and in the \ging\ observations of August 1988 (Dotani et
al., 1990).  They are also clearly evident in Mode 0 in observations
made by \sas\ in March 1979 (Basinska et al. 1980), by \hak\ in August
1979 (Kunieda et al. 1984a), and by \exo\ in 1985 (Stella et al.,
1988a; Lubin et al., 1993).  We observed with \rxtez\ a post-burst dip
on July 13, 1997 (Mode 0), but no pre-burst dip.  However, \tenma\
observed the RB in Mode 0 in August 1983, and did not detect any
preceding or following dips in 15 long ($>$100 seconds) type~II bursts
(Kunieda et al. 1984b).  Of course, instrument sensitivity is an
important factor in the detectability of dips in the PE, and \tenma\
may not have been sensitive enough to observe such dips.

There are some exceptions to this evolutionary pattern, however.  When
the RB was discovered in 1976, Mode I was observed, followed by Mode
II.  The RB then briefly returned to Mode I before ending in Mode II
(Lewin et al. 1976; Ulmer et al. 1977, Marshall et al. 1979).  In
April of 1977, \ariel\ and \sas\ saw the RB in Mode II, then Mode I,
and then finally Mode II again (White et al. 1978, Marshall et
al. 1979).

In addition, the duration of the sub-phases of Phase II can vary.  We
observed Mode 0 and Mode I to last for no more than $\sim$3 days each.
In 1979, however, \hak\ observed Mode 0 from August 8--16 (Kunieda et
al. 1984a).  In 1984, \tenma\ observed Mode 0 from July 2--5, and Mode
I from July 6--9 (Tawara et al. 1985).

There are also many idiosyncrasies associated with Phase II.  In
September 1977, when the RB was in Mode II, a larger than normal type
II burst followed a type~I burst on four occasions (Hoffman et
al. 1978a).  This led to the realization that the type~I bursts must
have come from the RB.  Also, Lubin et al. (1993) reported
``glitches'' that were observed following 10 of 84 long type~II bursts
observed with \exo\ in August 1985.

\section{Discussion and Summary}
\label{sec:discuss}

In four RB outbursts observed with \rxtez\, we have noted a more or
less consistent evolutionary pattern that is also reasonably
consistent (though not exclusively) with previous observations of the
RB.  The outbursts begin suddenly and rise to their peak persistent
emission (PE) level within three days.  Type~I bursts dominate for
15--20 days, with a steadily declining level of PE (Phase I).  By
approximately day 18 of the outbursts, type~II bursts appear (Phase
II).  There are three main types of type~II burst patterns: long,
flat-topped bursts; a series of short bursts followed by a large
burst; and rapid bursts at regular intervals.  The rapid, regular
bursts are the final phase of any outburst.

During the three outbursts in which Phase II was observed, the phase
did not begin until the PE luminosity had decreased to
$\sim$2$\times$10$^{37}$ erg s$^{-1}$.  It seems likely that the PE
level has important implications for the burst behavior of the RB.  

In addition, the fact that type~I bursts observed during Mode I occur
preferentially during the period of enhanced PE has important
implications for the mechanism that triggers type~I bursts.  Since the
periods of enhanced PE always follow a large type~II burst, it seems
plausible that the thermonuclear flash is triggered by the extra
amount of material that has been accreted onto the neutron star.

Using the relation $E$=$\beta (\frac{\Delta t}{100 s})^{\alpha}$ to
relate the energy of a type~II burst, $E$, to the waiting time to the
next burst, $\Delta t$, we find $\alpha$ values in the range
0.43--0.94 for Mode II and the large type~II bursts of Mode I.  The
short type~II bursts of Mode I have a relatively constant burst
separation, $\Delta t$, of $\sim$10 seconds, even though they vary in
energy.

The current average recurrence time for RB outbursts is $\sim$218
days.  This agrees reasonably well with all observed outbursts after
1983.  For outbursts prior to 1984, an average of $\sim$180 days seems
more suitable.

The possible reflares that might be present in two out of four
outbursts observed with \rxtez\ are reminiscent of the echoes observed
in some soft X-ray transients.  This could be the result of an echo of
the main outburst, in which either the companion star or the disk
itself is heated by X-rays from the main outburst.  The increased mass
flow to the accretion disk that results from the heating of the
companion star, or the excitation of the accretion disk if the disk
itself is heated, could produce a response that we observe as a
reflare of X-ray emission (Augusteijn, Kuulkers, \& Shaham 1993;
Tanaka \& Lewin 1995).  The delay between the onset of the outburst
and the echo, in this model, is related to the time required for
matter to be transferred from the location in the disk where the disk
instability occurs to the surface of the neutron star.

\subsection{Acknowledgments}
WHGL gratefully acknowledges support from the National Aeronautics and
Space Administration.  MK gratefully acknowledges the Visiting Miller
Professor Program of the Miller Institute for Basic Research in
Science (UCB).  We thank the Netherlands Organization for Research in
Astronomy ASTRON for financial support.

\end{document}